\documentclass[english,aps,prl,showpacs,superscriptaddress,floats,amsmath,amssymb,floatfix,twocolumn,nobalancelastpage]{revtex4-1}
\usepackage[T1]{fontenc}
\usepackage[latin9]{inputenc}
\usepackage{color}
\usepackage{textcomp}
\usepackage{amsmath}
\usepackage{amssymb}
\usepackage{graphicx}
\usepackage{esint}
\usepackage{wasysym}
\usepackage{lipsum}
\usepackage{hyperref}
\usepackage{babel}
\def\Xint#1{\mathchoice
   {\XXint\displaystyle\textstyle{#1}}%
   {\XXint\textstyle\scriptstyle{#1}}%
   {\XXint\scriptstyle\scriptscriptstyle{#1}}%
   {\XXint\scriptscriptstyle\scriptscriptstyle{#1}}%
   \!\int}
\def\XXint#1#2#3{{\setbox0=\hbox{$#1{#2#3}{\int}$}
     \vcenter{\hbox{$#2#3$}}\kern-.5\wd0}}

\def\dashint{\Xint-}
\begin{document}

\title{Thermal transport in a two-dimensional $\mathbb{Z}_2$ spin liquid} 

\author{Alexandros Metavitsiadis}

\email{a.metavitsiadis@tu-bs.de}

\selectlanguage{english}%

\affiliation{Institute for Theoretical Physics, Technical University Braunschweig,
D-38106 Braunschweig, Germany}

\author{Angelo Pidatella}

\email{angelo.pidatella@tu-dresden.de}

\selectlanguage{english}%

\affiliation{Institute for Theoretical Physics, Technical University Dresden,
D-01062 Dresden, Germany}

\author{Wolfram Brenig}

\email{w.brenig@tu-bs.de}

\selectlanguage{english}%

\affiliation{Institute for Theoretical Physics, Technical University Braunschweig,
D-38106 Braunschweig, Germany}

\date{\today}

\begin{abstract}
We study the dynamical thermal conductivity of the two-dimensional 
Kitaev spin-model on the honeycomb lattice. We find a strongly temperature
dependent low-frequency spectral intensity as a direct consequence 
of fractionalization of spins into mobile Majorana matter and a static 
$\mathbb{Z}_{2}$ gauge field. The latter acts as an emergent thermally activated 
disorder, leading to the appearance of a pseudogap which 
partially closes in the thermodynamic limit, indicating a dissipative 
heat conductor. Our analysis is based on complementary calculations of the 
current correlation function, comprising exact diagonalization by means of 
a complete summation over all gauge sectors, as well as a phenomenological 
mean-field treatment of thermal gauge fluctuations, valid at intermediate 
and high temperatures. The results will also be contrasted against the 
conductivity discarding gauge fluctuations.
\end{abstract}

\maketitle

\section{Introduction}

Thermal transport is an important tool to study elementary magnetic
excitations in local moment materials. This has been demonstrated in 
a large variety of systems displaying excitations, which range from 
conventional spin waves to exotic fractional
quasiparticles, including magnons \cite{Rives1969, Gorter1969,Lang1977, 
Nakamura1991,Hess2003,Hofmann2003,Sun2003,Yan2003}, 
triplons  \cite{Sologubenko2000,Hess2001}, spinons 
\cite{Sologubenko2000a,Sologubenko2001,Hess2007,Takayuki2008,Hlubek2010, 
Yamashita2010}, and emergent magnetic 
monopoles \cite{Kolland2012,Toews2013,Li2015,Tokiwa2016}.  
Most recently, the first thermal transport measurements have appeared 
in systems with strong spin-orbit coupling (SOC), which are potentially 
proximate to 2D spin-liquid states \cite{Leahy2017,Hirobe2016,Hentrich2017}. 

Quantum magnets with SOC have attracted considerable interest, because
they allow for directionally dependent highly anisotropic super-exchange,
which can lead to strongly frustrated quantum magnets \cite{Khaliullin2005, 
Jackeli2009,Chaloupka2010,Nussinov2015}. 
Among them is Kitaev's model on the honeycomb lattice \cite{Kitaev2006}.
It constitutes the rare case of a 2D spin system with an exactly known
spin-liquid ground state and fractionalization of spins in terms of
bulk Majorana fermions and $\mathbb{Z}_{2}$ gauge fields \cite{Kitaev2006, 
Feng2007,Chen2008,Nussinov2009,Mandal2012}.
Part of its quantum phases are perturbatively equivalent to the toric
code \cite{Kitaev2003}, providing a direct link to paradigmatic models
of topological order \cite{Wen1989,Wen1990}. In finite magnetic fields
the $\mathbb{Z}_{2}$ vortices acquire non-abelian anyonic statistics and the
Majorana Dirac spectrum opens a gap displaying a chiral edge mode
\cite{Kitaev2006}.

There is an ongoing quest for Kitaev materials with 2D honeycomb variants
$\mathrm{Na_2IrO_3}$ \cite{Singh2010}, $\alpha$-$\mathrm{Li_{2}IrO_{3}}$  
\cite{Singh2012}, $\alpha$-$\mathrm{RuCl_{3}}$ \cite{Plumb2014}, and 3D polymorphs 
$\beta$-,$\gamma$-$\mathrm{Li_{2}IrO_{3}}$ \cite{Modic2014,Takayama2015}, as well
as triangular lattice versions $\mathrm{Ba_{3}IrTi_{2}O_{9}}$ under scrutiny 
\cite{Dey2012}. Presently, all compounds show significant non-Kitaev
exchange. The role of coupling to extrinsic degrees of freedom,
such as phonons, is an open issue \cite{Hentrich2017}. In pursuit
of signatures of fractionalization, an enormous amount of research has
been performed on the spin dynamics in Kitaev models and materials,
including the dynamic structure factor \cite{Banerjee2016,Banerjee2016a, 
Gohlke2017}, Raman scattering \cite{Sandilands2015,Nasu2016}, and nuclear 
magnetic resonance \cite{Baek2017}. Thermal conductivity measurements on 
$\alpha$-RuCl$_{3}$ \cite{Leahy2017,Hirobe2016,Hentrich2017} have
yet been confined to the longitudinal component $\kappa_{xx}$ and
reveal \cite{Hentrich2017} that the heat transport seems intimately
related to spin-phonon coupling. Observations of putative chiral edge
modes, using off-diagonal $\kappa_{xy}$ at finite magnetic fields
are still lacking.

Theoretically, fractionalization has long been a topic for magnetic transport in
1D quantum magnets, due to the existence of spinons in Heisenberg chains and
dimerized or frustrated variants thereof \cite{FHM2007, Hess2007a}. One key
question is the dissipation of currents, which has been investigated extensively
at zero frequency (dc) and momentum in connection with the linear response Drude
weight (DW) \cite{Shastry1990,Zotos1997,Klumper2002,Fujimoto2003,FHM2003,
Prosen2011,Steinigeweg2014}, which is the non-dissipating dc part of the current
autocorrelation function and, if existent, indicates a ballistic channel of the
fractional quasiparticles.

First theoretical studies, of heat transport in Kitaev models have
been carried out on chains \cite{Steinigeweg2016} and ladders \cite{Metavitsiadis2016}
with very different conclusions. In the former, gauge fields are absent
and the chain is found to be a perfect, ballistic heat conductor with
a finite thermal DW. The ladder is the simplest quasi-1D descendant
of the honeycomb lattice model featuring both, matter fermions and
$\mathbb{Z}_{2}$ gauge fields. It is found to display no ballistic channel
and a zero frequency insulating pseudogap. This is a direct consequence
of fractionalization, with the static gauge fields acting as an emergent,
thermally induced disorder, which scatters the current carrying mobile
Majorana matter. Since dimensionality of the Majorana matter could
have a significant impact on the scattering from the gauge field,
the prime motivation of the present work is to extend the ideas of
Ref.~[\citenum{Metavitsiadis2016}] to 2D. As our central results, we find
that similar to Ref.~[\citenum{Metavitsiadis2016}], ballistic channels
are suppressed and finite low-frequency spectral weight is generated in
the dynamical conductivity by scattering from the gauge field. However,
in sharp contrast to the ladder, the pseudogap does not fully survive
the thermodynamic limit in 2D, leading to a dissipative heat conductor,
rather than an insulator.

The paper is organized as follows. In Section \ref{secKitaev},  
we provide details of the Kitaev model as needed for this
work. In Section \ref{linResp} we summarize magnetic heat transport theory
in the linear response regime. In Section \ref{results} we present and
compare our results, derived from three complementary methods, i.e. exact
diagonalization (ED), Sec.~\ref{secED}, average gauge
configuration (AGC) calculations, Sec.~\ref{secAGC}, and zero vortex
sector (ZVS) analytic evaluations, Sec.~\ref{secUni}. Lastly,  
Sec.~\ref{secConcl} contains our conclusions.

\section{Kitaev model}\label{secKitaev}

\begin{figure}[tb]
\centering{}\includegraphics[width=0.95\columnwidth]{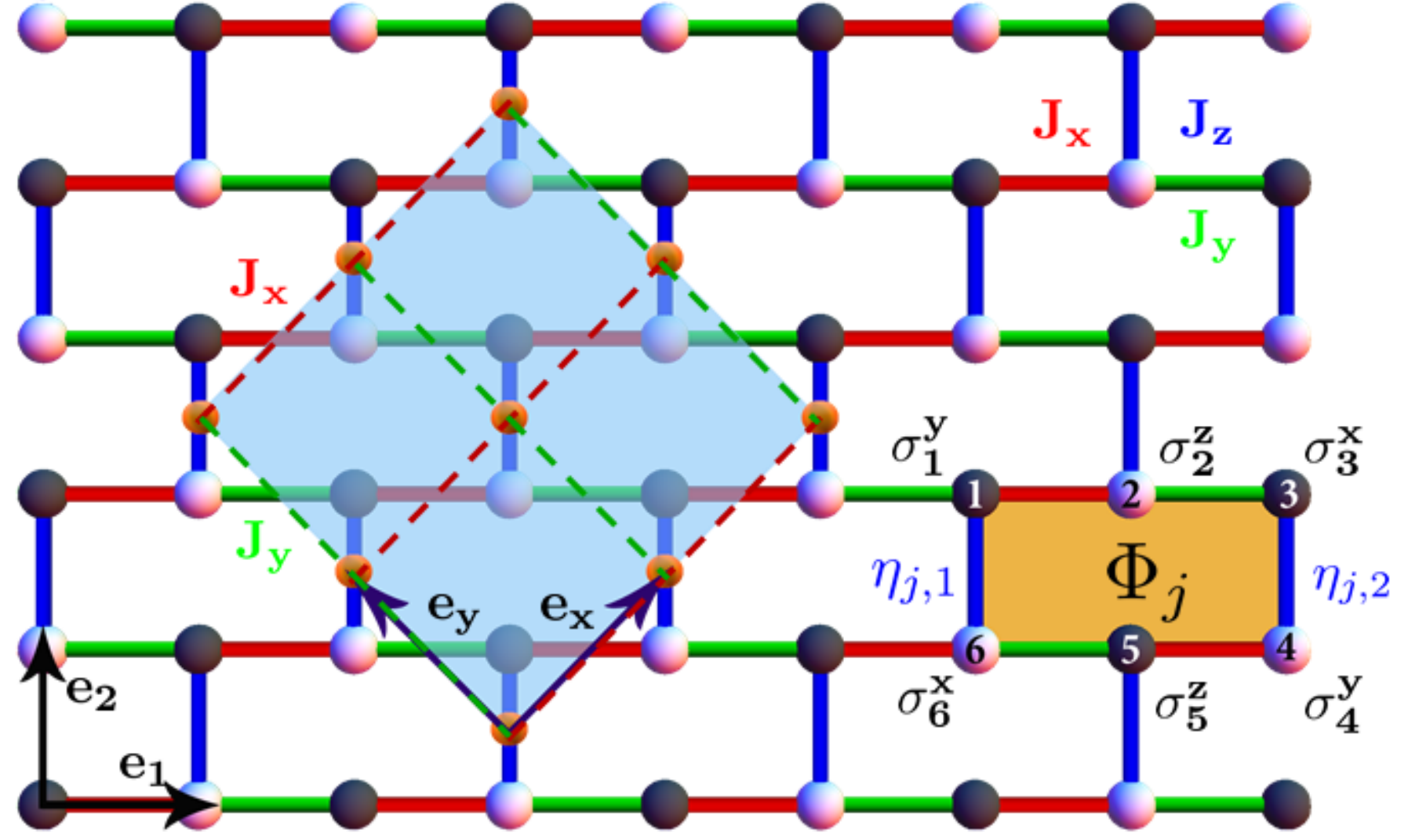}
\caption{\label{fig:a1} Kitaev model on a honeycomb lattice, 
deformed into the  so-called brickwall lattice. 
The $J_{x}$, $J_{y}$, and $J_{z}$ exchange links are indicated by red, green, 
and blue lines respectively, and the two sublattices with dark and light color 
bullets. The lattice is formed along the $\mathbf{e}_1$, $\mathbf{e}_2$ 
directions. The dark yellowish bullets lying on the middle of the $z$-links, 
indicate the vertices of the effective square lattice (ESL), rotated to the 
left by 45\textdegree, with unit vectors $\mathbf{e}_x$, $\mathbf{e}_y$ 
(light blue shaded region). The exchange interactions for the ESL are 
$J_x$ and $J_y$ along the $\mathbf{e}_x$ and $\mathbf{e}_y$ directions 
respectively. The flux operator $\Phi_j$ of the $j$-th plaquette, is the 
product of the 6 spin operators around the plaquette, as shown in the dark 
yellowish highlighted region. The eigenvalue of the flux operator on that 
plaquette is equal to the product of the two corresponding $\eta$ fields 
belonging to the same plaquette, i.e., 
$\Phi_j = \prod_{l=1}^6 \sigma_{jl}^{\alpha(l)}=\eta_{j,1} \eta_{j,2}$.
}
\end{figure}

In this section we briefly summarize several points, clarifying the use of
Kitaev's model \cite{Kitaev2006} in this work. The Hamiltonian reads
\begin{equation}
H=\sum_{\langle l,m\rangle,\alpha(\langle l,m\rangle)}J_{\alpha}\sigma_{l}^{\alpha}
\sigma_{m}^{\alpha}\,,\label{eq:a1}
\end{equation}
where $ \langle l,m \rangle$ refer to the sites on the nearest neighbor bonds 
of the honeycomb lattice. For simplicity, in this work, we will envisage 
the honeycomb lattice to be deformed into the so-called brickwall lattice 
(BWL) \cite{Feng2007,Chen2008}, shown in Fig.~\ref{fig:a1}.  $J_{ \alpha}$ 
and $\sigma^{\alpha}$ are the exchange coupling constants and Pauli matrices 
respectively for coordinates $\alpha=x,y,z$. The relation 
$\alpha( \langle l,m \rangle)$ can be read off from the red, green, and blue  
bond coloring in Fig.~\ref{fig:a1}. It is known that this 
model can be mapped onto free spinless (`matter') fermions in the presence of 
static $\mathbb{Z}_{2}$ gauge fluxes. The allowed values $ \pm1$ of the latter are 
related to the eigenvalues of the conserved operator 
$ \Phi= \prod_{l=1 \dots6} \sigma_{l}^{\alpha(l)}$ around each plaquette, 
Fig.~\ref{fig:a1}, where $ \alpha(l)=x,y,z$, refers to that component of the 
exchange link which is \emph{not} part of the loop passing site 
$l$ \cite{Kitaev2006}. For the remainder of this work we set
$\hbar$, $k_B$, and the lattice constant $a$ to unity, and choose $J_z$ as
the unit of energy.    

Several routes have been established to map the spins in Eq. (\ref{eq:a1}) to 
fermions, e.g.~using overcomplete sets of Majorana fermions \cite{Kitaev2006},
Jordan-Wigner transformation \cite{Feng2007,Chen2008}, or 
bond algebras \cite{Nussinov2009}.  The resulting final Hamiltonian 
reads \cite{Chen2008}
\begin{equation}
H=\sum_{\mathbf{r}}h(\mathbf{r})\, 
,\label{eq:a2}
\end{equation}
with $h(\mathbf{r})$ the single particle local energy   
\begin{eqnarray}
h(\mathbf{r}) & = & J_{x}(d_{\mathbf{r}}^{\dagger}+
d_{\mathbf{r}}^{\phantom{\dagger}})(d_{\mathbf{r}+\mathbf{e}_{x}}^{\dagger}
-d_{\mathbf{r}+\mathbf{e}_{x}}^{\phantom{\dagger}})+J_{y}
(d_{\mathbf{r}}^{\dagger}+d_{\mathbf{r}}^{\phantom{\dagger}})\nonumber \\
 &  & \times(d_{\mathbf{r}+\mathbf{e}_{y}}^{\dagger}-d_{\mathbf{r}+
\mathbf{e}_{y}}^{\phantom{\dagger}})+J_{z}\eta_{\mathbf{r}}(
2d_{\mathbf{r}}^{\dagger}d_{\mathbf{r}}^{\phantom{\dagger}}-1)\, . 
\hphantom{.aaaa}\label{eq:a3}
\end{eqnarray}
$d_{ \mathbf{r}}^{(\dagger)}$ and $ \eta_{ \mathbf{r}}= \pm1$ refer to the 
spinless matter fermions and the gauge fields, which can be visualized to be 
located on the sites of a dual lattice of the $z$-bonds of the BWL, forming 
an effective square lattice (ESL), Fig.~\ref{fig:a1}.   
In the fermionic representation 
$\Phi_{ \mathbf{r}}= \eta_{ \mathbf{r}} \eta_{ \mathbf{r}+
\mathbf{e}_{1}}$, for the brickwall lattice, and 
$\Phi_{ \mathbf{r}}= \eta_{ \mathbf{r}} \eta_{ \mathbf{r}+\mathbf{e}_{x} 
-\mathbf{e}_{y}}$ for the ESL. From the preceding, it is rather apparent that 
the ESL model lacks $C_4$ symmetry and the two diagonal directions 
$\mathbf{e}_{x} \pm \mathbf{e}_{y}$ are distinct. 

While the focus of our work is on bulk transport properties, we state three
remarks of caution regarding boundary conditions. First, the mapping from
Eq.~(\ref{eq:a1}) is exact only if periodic boundary conditions (PBCs) are used
along the $\mathbf{e}_2$ direction of the BWL \cite{Feng2007,Chen2008}.
Requiring PBCs also along the $x,y$-chains ($\mathbf{e}_1$ direction of the BWL),
requires consideration of surface terms \cite{Chen2008,Mandal2012}, as known
from any Jordan-Wigner type of mapping. It has been shown recently, that for
bulk thermal transport on Kitaev ladders such surface terms have no relevant
effect \cite{Metavitsiadis2016}, and therefore we discard them.  Second, to
describe bulk properties based on the ESL, it is natural to apply PBCs along the
$\mathbf{e}_{x,y}$ directions of the lattice. This implies nonstandard $O(1/L)$
finite size corrections for a system of $N=L \times L$ sites. We do not expect
these to be of any qualitative relevance. Third, and finally, the spectrum of
the Kitaev model is highly degenerate. The relevance of this for the current
correlation function is briefly commented on in App.~\ref{appDeg}.

\section{Thermal transport}\label{linResp}

The primary goal of this work is to evaluate the dynamical equilibrium 
\emph{bulk} thermal conductivity of the Kitaev model. In this section, 
and for completeness, we recollect the basic ingredients for this. 

To start, linear response theory with respect to a real space dependent 
local equilibrium temperature $T+ \delta T( \mathbf{r})$ has to be 
performed, based on a canonical density matrix $\rho=\exp[
- \int d^{3}r \,( \beta + \delta \beta(r)) h(\mathbf{r})]$, where 
$ \beta=1/T$, and $h( \mathbf{r})$ is a local energy density which 
has to fulfill $H= \int d^{3}r\,h(\mathbf{r})$, where $H$ is the Hamiltonian. 
In this framework, the linearized expectation value of the $\mu$ component 
of the energy current ${\cal J}_\mu$ in $d$ dimensions  is obtained from 
the dynamical thermal conductivity tensor $ \kappa_{\mu\nu}(\mathbf{q},\omega)$ 
at wave vector $\mathbf{q}$ and frequency $\omega$ through 
\begin{equation}
\langle{\cal J}_{\mu}(\mathbf{q},\omega)\rangle=\sum_{\nu=1}^{d}
\kappa_{\mu\nu}(\mathbf{q},\omega)\partial_{\nu}T(\mathbf{q},\omega)\,. 
\label{eq:a4}
\end{equation}
The equilibrium thermal expectation value $\langle A\rangle$ reads $\langle
A\rangle = \mathrm{Tr} A e^{-\beta H}/Z$, with $Z=\mathrm{Tr}e^{-\beta H}$ 
the partition function.  The spectrum $\kappa_{ \mu \nu}^{\prime}
(\omega)$, namely the real part of the thermal conductivity, follows from the
Fourier transform of the current correlation function,
$C_{\mu\nu}(\omega)$,
\begin{equation} 
\kappa_{\mu\nu}^{\prime}(\omega)  =  \frac{\beta}{2\omega} \left(
1-e^{-\beta\omega}\right)\, C_{\mu\nu}(\omega)~,  \label{eq:a6} 
\end{equation}
\begin{equation} 
C_{\mu\nu}(\omega)  = \int dt e^{i\omega t} C_{\mu\nu}(t) ~, ~~ 
C_{\mu\nu}(t) = \frac{1}{N}\ \langle{\cal J}_{\mu}(t){\cal J}_{\nu}\rangle. \label{eq:b6} 
\end{equation}
It is customary to decompose $C_{\mu\nu}(\omega)$ as \cite{FHM2007} 
\begin{equation}
C_{\mu\nu}(\omega)=4\pi D_{\mu\nu}T^{2}\,\delta(\omega)+C_{\mu\nu}^{reg}(\omega)
\,,\label{eq:a7}
\end{equation}
where the \emph{regular} part refers to $C_{ \mu \nu}^{reg}( \omega)= C_{
\mu \nu}(\omega \neq0)$ and the \emph{Drude} weight (DW) $D_{ \mu \nu}$ is a
measure for the ballistic contribution to the heat flow
\begin{equation}
D_{\mu\nu}=\frac{\beta^{2}}{2ZV}\sum_{E_{l}=E_{m}}e^{-\beta E_{l}}\langle l|
{\cal J}_{\mu}|m\rangle\langle m|{\cal J}_{\nu}|l\rangle\,.\label{eq:a8}
\end{equation}
Because of Eq.~(\ref{eq:a6}), 
$ \kappa_{\mu \nu}^{ \prime}( \omega)= 2\pi D_{ \mu \nu} \delta( \omega) + 
\kappa_{\mu\nu}^{reg}(\omega)$.  
Whenever $D_{ \mu \nu} \neq0$, the system 
is a perfect heat conductor in channel $ \mu \nu$. Otherwise it is a 
dissipative conductor with a limiting dc heat conductivity of 
$\kappa{}_{ \mu \nu}^{dc} = 
\kappa{}_{ \mu\nu}^{\prime}(\omega \rightarrow0)$.  If both, $D_{ \mu \nu}=0$ 
and $ \kappa{}_{\mu\nu}^{dc}=0$, the system is an insulator in channel 
$ \mu \nu$. 

To determine the energy current $\boldsymbol{\cal J}$ we turn to a real
space version of the continuity equation $\partial_{t}h( \mathbf{r})+
\nabla\cdot \boldsymbol{{ \cal J}}(\mathbf{r})=0$, which is more amenable
to describe the spinless fermions of Hamiltonian (\ref{eq:a2}) and
(\ref{eq:a3}), which comprise a real space dependent potential by virtue of
$\eta_{\bf r}$. To this end we use the polarization operator
$\mathbf{P}$ \cite{Metavitsiadis2016},
\begin{equation} 
\boldsymbol{{\cal J}}  =  i[H,\mathbf{P}]~, \quad\text{with}\quad 
\mathbf{P}  =  \sum_{\mathbf{r}}\mathbf{r}\,
h(\mathbf{r})\,,\label{eq:a9}
\end{equation}
which yields the same current operator as the continuity equation in the limit of
$\mathbf{q}\rightarrow0$ for a homogeneous system. For the Kitaev
model on the ESL and using our \emph{definition} of the energy density, given in
Eq.~\eqref{eq:a3}, we arrive at the energy current
\begin{equation}
{\cal J}_{\mu} =2iJ_\mu \sum_{\mathbf{r}}
[ J_{z} \eta_{\mathbf{r}}b_{\mathbf{r}}b_{\mathbf{r}-\mathbf{e}_{\mu}}
+ \tau_\mu J_{\bar{\mu}}b_{\mathbf{r}}b_{\mathbf{r}+\mathbf{e}_{x}-\mathbf{e}_{y}}]~, 
\label{eq:a10}
\end{equation}
where $b_{ \mathbf{r}}=(d_{ \mathbf{r}}^{ \dagger}+d_{ \mathbf{r}})$, 
$\bar \mu = y(x)$ and $\tau_\mu=+(-)$ for $\mu=x(y)$. 
From the expression above, one can readily see that also the energy current 
operator is diagonal in the gauge fields. 

We caution that the only requirement for $h(\mathbf{r})$ is, that $H=\int
d^{3}r\,h(\mathbf{r})$. This may be a reason for differing
\emph{quantitative} results for the Drude weight and the regular
conductivity spectrum, obtained in recent studies of various frustrated and
spin ladder models \cite{Alvarez2002, FHM2003, Zotos2004, Steinigeweg2016}. 
However, it is generally believed that universal
\emph{qualitative} statements, concerning the existence or absence of
finite Drude weights and dc conductivities are insensitive to the freedom
of choice for the energy density.

\section{Evaluation of heat current correlation functions}\label{results}

Even though Kitaev's model comprises free fermions, the distribution of the
$\eta_{\mathbf{r}}$ in real space, renders analytical evaluation of thermal
traces infeasible. Numerically, quantum Monte-Carlo (QMC) methods have been
used for a variety of observables \cite{Motome2014a,
Motome2015b,Motome2015a}.  Regarding thermal transport, to the best of our
knowledge, exact diagonalization (ED) \cite{Metavitsiadis2016}, summing over
all gauge configurations, supplemented also by approximate methods, has
been used first to evaluate $C_{\mu\mu}(t)$ for the Kitaev Hamiltonian on a
ladder \cite{Metavitsiadis2016}. Here we will extend this work to d=2
dimensions analyzing the \emph{longitudinal} heat transport properties,
i.e.~$C_{\mu\mu}$, $\kappa'_{\mu\mu}$, and $D_{\mu\mu}$, at finite
temperatures, using unbiased exact diagonalization, as well as an
approximate ensemble of average gauge configurations. In addition, we
perform an illustrative evaluation of $C_{\mu\mu}(\omega)$, and
$D_{\mu\mu}(T)$ based on only the uniform gauge. In following, we focus 
on the isotropic gapless point $J_x=J_y=J_z$ unless mentioned otherwise.

\subsection{Exact diagonalization (ED)}\label{secED}

Since the energy current operator is diagonal in the gauge field, the 
correlation function $C_{ \mu \mu}(t)$  can be written as
\begin{equation}
C_{\mu\mu}(t)=\frac{1}{ZN}Tr_{\eta}[Z_{d(\eta)}\,\langle{\cal J}_{\mu}(t)
{\cal J}_{\mu}\rangle_{d(\eta)}]\,,\label{eq:a11}
\end{equation}
where $N$ is the number of lattice sites, the subscript $d(\eta)$ refers 
to tracing over matter fermions at a \emph{fixed} gauge field state and the 
subsectors' partition functions $Z_{d(\eta)}$ sum up to the total partition 
function  $Z$. To numerically evaluate Eq.~\eqref{eq:a11},  we resort to ED. 
To this end, we define a $2N$ component operator $\mathbf{D}^{ \dagger}=
(d_{1}^{ \dagger}, \ldots, d_{N}^{\dagger},
d_{1}^{ \phantom{ \dagger}},\ldots,  
d_{N}^{\phantom{ \dagger}})$ of the matter fermions. The indices 
$\{1,2, \ldots, N \}$ label all sites $ \mathbf{r}$ of the ESL. 
In terms of $ \mathbf{D}^{ \dagger}$ the Hamiltonian 
and the current are set up in real space as 
$H= \mathbf{D}^{ \dagger} \mathbf{h}( \eta)\mathbf{D}$, and  
${ \cal J}_{ \mu}= \mathbf{D}^{ \dagger} \mathbf{j}_{ \mu}( \eta) \mathbf{D}$. 
Both $\mathbf{h}( \eta)$ and $\mathbf{j}_{ \mu}( \eta)$ 
are  $2N \times2N$ matrices, which depend 
on the actual state of the gauge field $\eta= (\eta_{1}, \eta_{2}, 
\ldots, \eta_{N})$.  For each given $ \eta$ we compute a Bogoliubov transformation 
$ \mathbf{U}$, which introduces canonical quasiparticle fermions 
$\mathbf{A}^{ \dagger}=(a_{1}^{ \dagger}, \ldots, a_{N}^{ \dagger},a_{1}^{
\phantom{ \dagger}}, \ldots, a_{N}^{ \phantom{ \dagger}})$ via $ \mathbf{A}=
\mathbf{U}^{ \dagger} \mathbf{D}$ and maps the Hamiltonian to 
$H= \frac{1}{2}\mathbf{A}^{\dagger} \mathbf{E} \mathbf{A}$, where 
$\mathbf{E}$ is diagonal and 
$\mathrm{diag}(E)=( \varepsilon_{1}, \ldots, \varepsilon_{N},- \varepsilon_{1}, 
 \ldots, -\varepsilon_{N})$, with $\varepsilon_j$ being the quasiparticle energies.  

With these definitions, the current correlation function in a fixed gauge 
configuration reads 
\begin{eqnarray}
C_{\mu\mu}^{\eta}(\omega) &= & \frac{2\pi}{N}\sum_{klmn}
L_{kl}L_{mn}\Big(\langle A_{k}^{\dagger}A_{n}^{\phantom{\dagger}}
\rangle\langle A_{l}^{\phantom{\dagger}}A_{m}^{\dagger}\rangle\nonumber \\
 & -& \langle A_{k}^{\dagger}A_{m}^{\dagger}\rangle\langle A_{l}^{\phantom{\dagger}}
A_{n}^{\phantom{\dagger}}\rangle\Big)\delta(\varepsilon_{l}-\varepsilon_{k}-\omega)
\,,\label{eq:9}
\end{eqnarray}
where $ \mathbf{L}= \mathbf{U}^{ \dagger} \mathbf{j}( \eta) \mathbf{U}$ and $ \langle
A_{ \mu}^{( \dagger)}A_{ \nu}^{( \dagger)} \rangle$ is either zero, $f_{j}$, or
$(1-f_{j})$, depending on the components of the spinor $ \mathbf{A}$ involved, and  
$f_j=1/(e^{\beta\varepsilon_j} +1)$ is the Fermi-Dirac distribution. 
Since the partition functions $Z_{d( \eta)}$ are byproducts of the ED for 
each gauge subsector, tracing the latter and Eq.~(\ref{eq:a9}), as in 
Eq.~(\ref{eq:a11}) completes the evaluation of $C_{ \mu \mu}( \omega)$.

\begin{figure}[tb]
\includegraphics[width=0.95\columnwidth]{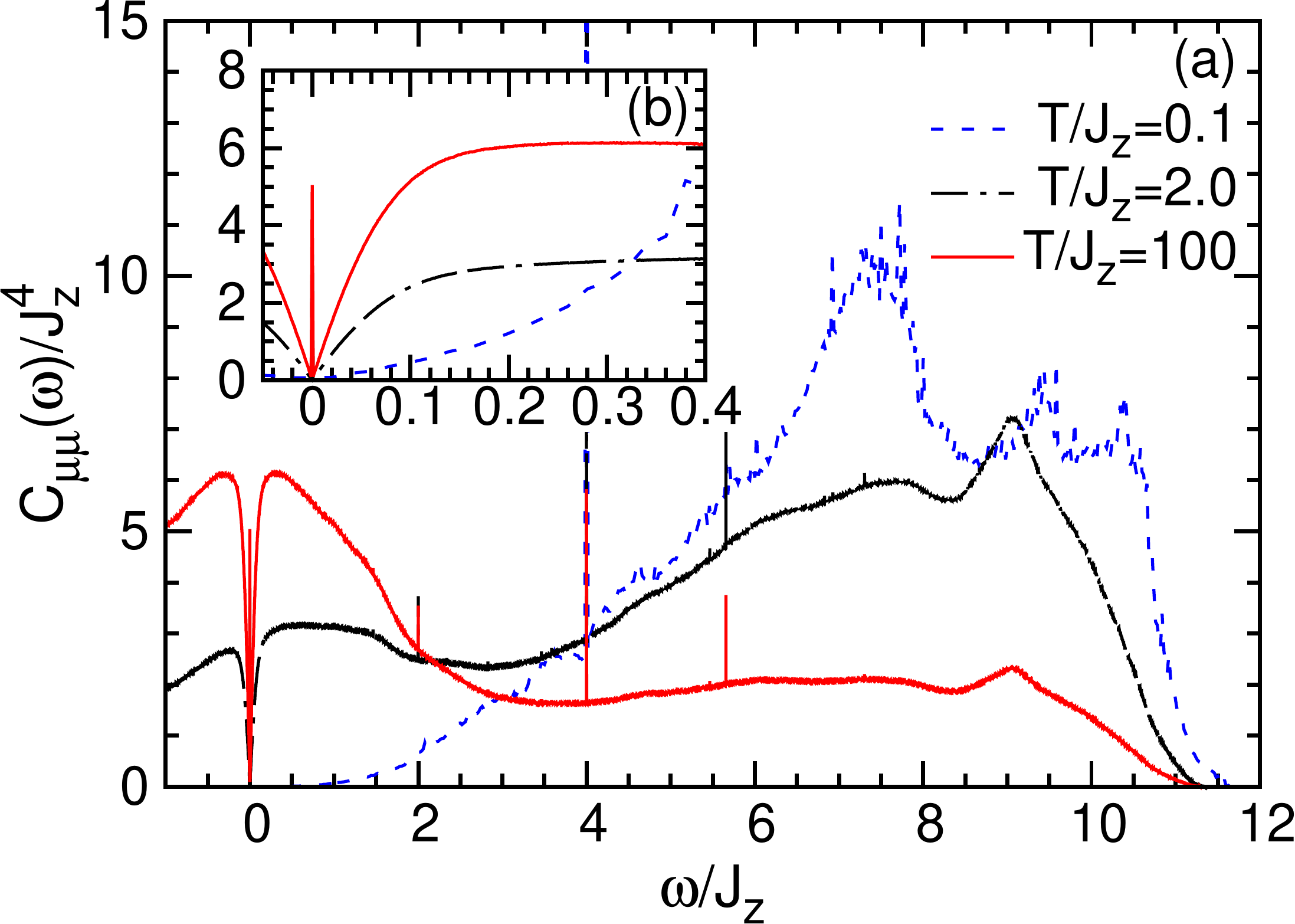}
\caption{(a) $C_{\mu\mu}(\omega)$ obtained by ED for a lattice of $N=6\times 6$ sites 
at three different temperatures $T/J_z=0.1,2,100$. (b) Low frequency zoom of (a). 
}\label{figCOmegaED}   
\end{figure}

In Fig.~\ref{figCOmegaED}, we present the frequency
dependence of the correlation function $C_{\mu\mu}(\omega)$ for three temperatures
$T/J_z=0.1,2,100$. We note that we focus only on the positive frequency spectrum, $
\omega>0$, because $C_{\mu\mu}(- \omega)=e^{- \beta \omega}C_{\mu\mu}(\omega)$, as
required by detailed balance. To reduce the computational effort for the evaluation
of full traces over the $2^N$ possible gauge field configurations, we make use of
translation symmetry.  This allows us to reach systems up to $N=36$ sites
corresponding to an enormous Hilbert space dimension of the underlying spin model of
$2^{72}$ states. The $\delta$ functions are binned in windows of $\delta
\omega=10^{-3}$, except for the lowest temperature, $T/J_z=0.1$, at which a binning
of $\delta \omega = 0.02$ has been chosen due to the larger finite size effects. The
sharp peaks at finite frequencies are amplified by the very fine binning, and they are 
not expected to survive in the thermodynamic limit.

For all temperatures $T/J_z \gtrsim 1$ the spectrum is {\it qualitatively}
different from that discarding gauge excitations, which will be analyzed in
Section \ref{secUni}, see Fig.~\ref{fig4}. We understand this central
result to be a clear indication of fractionalization, where the matter
fermions scatter off gauge field degrees of freedom, with the latter acting
as a thermally activated disorder. In fact, Eq.~(\ref{eq:9}) allows for two
types of spectral contributions, namely quasiparticle, i.e. $\varepsilon_l
\varepsilon_k > 0$, or pairbreaking, i.e. $\varepsilon_l \varepsilon_k < 0$
transport. The high frequency spectral weight ($\omega/J_z\gtrsim 6$)  
in Fig.~\ref{figCOmegaED} is solely generated by the pairbreaking terms, 
and it is only quantitatively affected by the gauge disorder. 
Contrarily, the quasiparticle contribution is related to the matter fermion density 
relaxation and therefore is strongly affected by scattering from the gauge 
fields. In the ground state gauge, the complete quasiparticle transport will 
accumulate into only a single $T$-dependent DW, see Sec.~\ref{secUni}. In the 
presence of the gauge excitation however, most of the DW spreads over a finite 
low-$\omega$ range, which displays an increasing weight in 
Fig.~\ref{figCOmegaED} as $T$ increases. The latter is due to the
temperature dependence of the matter fermion occupation number. We note
that on any finite systems remnants of a DW will remain within the
spectrum.

At very low temperatures, $T/J_z\lesssim 0.1$ the gauge excitations will
start to freeze out and the correlation function will approach the form of
$C_{\mu\mu}(\omega)$ within the ground state sector, Fig.~\ref{fig4}. This
regime suffers from large finite size effects and it is difficult to be
tackled with our methods (see also the discussion in the context of
Fig.~\ref{figThermodynamics}).

The low-$\omega$ spectral hump in the correlation function displays a
clearly visible, sharp dip as $\omega\rightarrow 0$, with
$C(\omega\rightarrow 0)\simeq 0$, as can be read off from
Fig.~\ref{figCOmegaED}(b). This is particularly obvious at elevated
temperatures. The behavior of this low frequency pseudogap with system
size is crucial in order to characterize the system as conducting or
insulating in the thermodynamic limit, and requires a careful finite size
analysis.  Either the pseudogap will close as $L\rightarrow\infty$ and the
system will have a conducting dc channel, or the pseudogap remains
open. In the latter case the system will be characterized by the
presence(absence) of a finite DW as an ideal conductor(insulator). These
issues relate the system directly to questions of disorder in
Dirac semi-metals \cite{NERSESYAN1995561, PhysRevLett.97.146805}.  
While our ED provides clear evidence for
signatures of fractionalization in the dynamical correlation function, a
convincing answer to the behavior of the pseudogap with system size
requires larger lattices, which we will tackle with the average gauge
configuration approach, presented in Sec.~\ref{secAGC}.

\begin{figure}[tb]
\includegraphics[width=0.95\columnwidth]{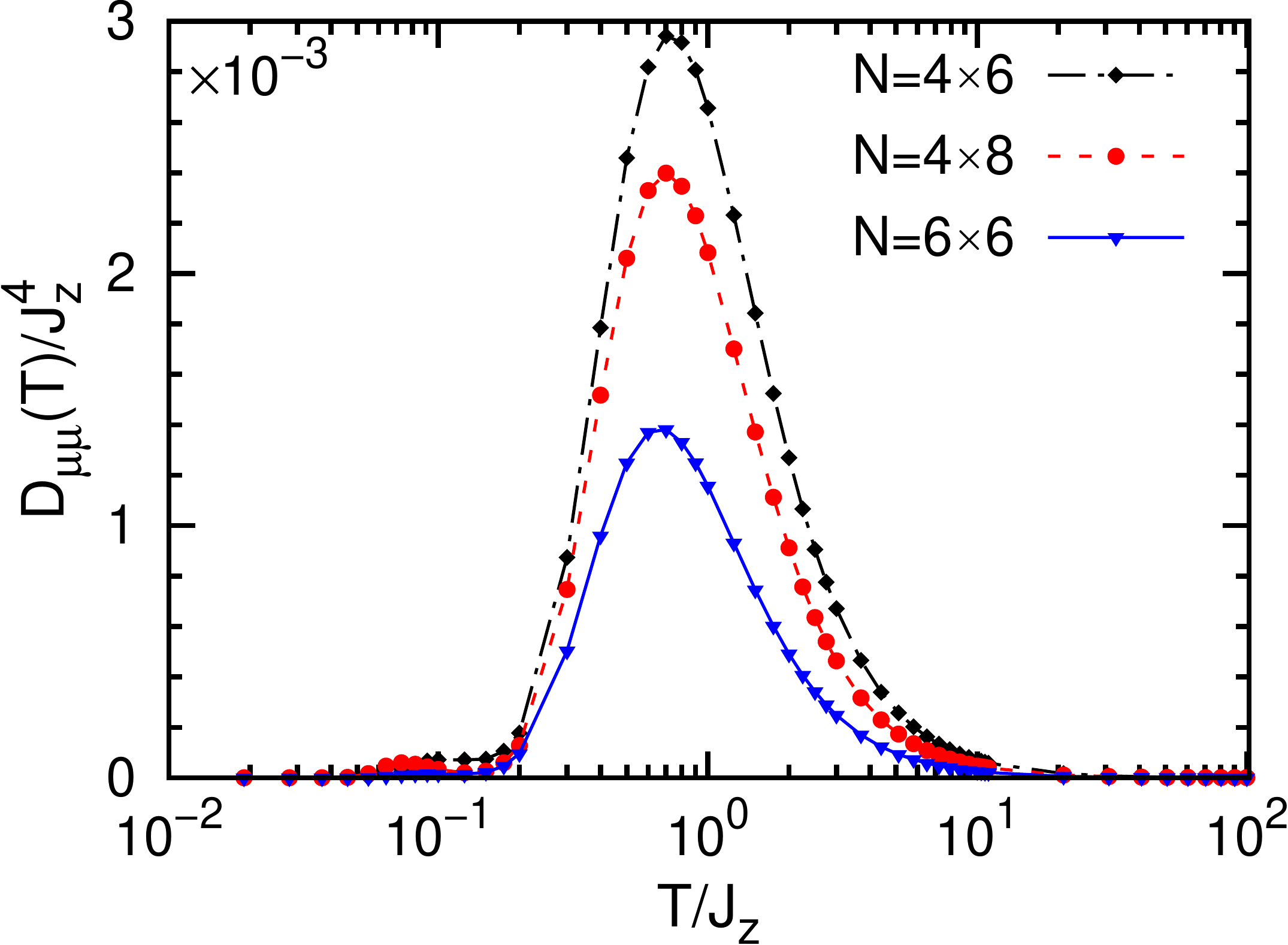}
\caption{Drude weight $D_{\mu\mu}(T)$ versus temperature for three different system sizes 
obtained via ED.}\label{figDW}  
\end{figure}

Next, we focus on the temperature dependence of the ballistic contribution
to the thermal conductivity, namely the DW,
Eqs.~\eqref{eq:a7}, and \eqref{eq:a8}. In Fig.~\ref{figDW}, we present the
temperature dependence of the DW for three different system sizes, acquired
from the degeneracy plateau  as usual \cite{FHM2007}.  The general form of
the temperature dependence of the DW is that of a typical spin system
exhibiting a maximum around $T\approx J_z$. With increasing system size, the
magnitude of the DW is reduced.  Note that this is different from the
behavior of the DW obtained for other transport quantities in 
Ref.~[\citenum{Motome2015b}]. Although the system sizes at hand do not allow
for a safe finite size extrapolation, our findings are suggestive of a
vanishing DW in the thermodynamic limit. This picture is further
corroborated by the AGC method, presented in Sec.~\ref{secAGC} \cite{refAppCom}. 

\subsection{Average gauge configuration (AGC)}\label{secAGC}

In this section, we introduce an {\it approximate} method to evaluate the
current correlation function, capturing the main physics, and allowing to
reach systems of $\sim 60\times 60$ sites, i.e. $\sim O(100)$ larger than
with ED, which is crucial to understand the low frequency regime of the
correlation function. The main idea is to reduce the full trace
$Tr_{\eta}[\dots]$ to an average $\langle \ldots \rangle_{n(T)}$ over only
dominant gauge configurations, set by a temperature dependent \emph{mean}
density $n(T)$ of elementary gauge excitations off the gauge ground state.
I.e., we reduce the evaluation of $C_{\mu\mu}(t)$ to a \emph{disorder
problem} in a system of free fermions with an emergent temperature
dependent defect density
\begin{equation}
C_{\mu\mu}(t)\approx\langle\langle{\cal J}_{\mu}(t){\cal J}_{\mu}
\rangle_{d(\eta)}\rangle_{n(T)}\,.\label{eq:11}
\end{equation}
Several comments are in order for this approach. First, while the
Hamiltonian (\ref{eq:a2},\ref{eq:a3}) is formulated in terms of matter
fermions and gauge fields $\eta_{\bf r}$, the physical degrees of freedom
are rather fermions and fluxes. In turn, depending on the temperature,
fluctuations in $n(T)$ may be very large, rendering a mean field treatment
in terms of the number of excited fluxes $\Phi(T)$ more appropriate. On the
Kitaev ladder \cite{Metavitsiadis2016}, this can be achieved by a direct
mapping between $n(T)$ and $\Phi(T)$. On the honeycomb lattice this is not
feasible. To make progress, we confine ourselves to temperatures above
a scale $T_R$, which is elevated enough, such that a large number of fluxes is
excited. Then, random gauge and flux ensembles will both behave similarly.

\begin{figure}[tb]
\centering 
\includegraphics[width=0.95\columnwidth]{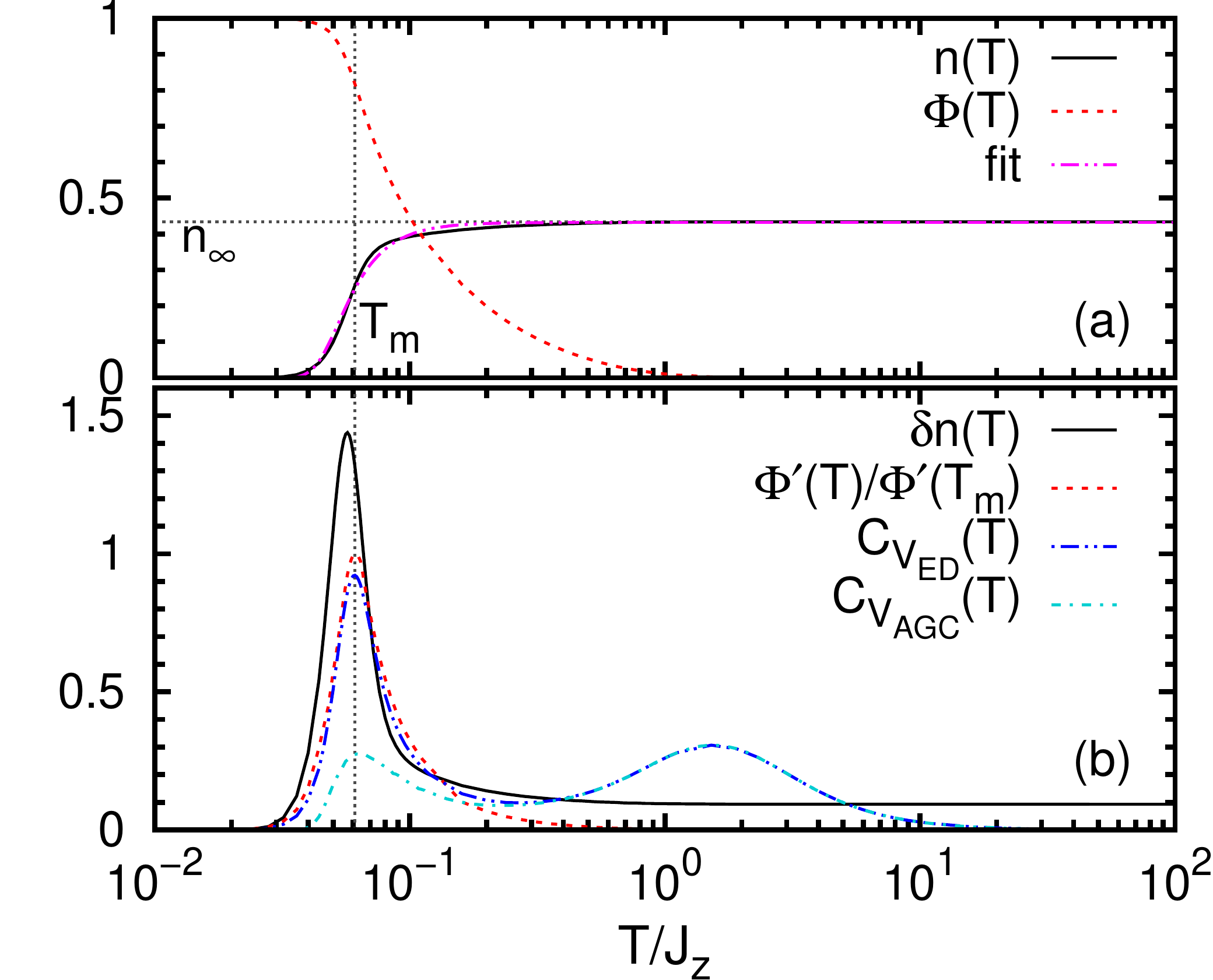}
\caption{(a) Temperature dependence of the mean density of gauge excitations
$n(T)$, and the flux density $\Phi(T)$. A fit, $n(T)
\sim 1/[e^{(\Delta/T)^b}+1]$ is shown, with $\Delta=0.06$, and $b=3.5$. 
The infinite temperature limit of $n(T\rightarrow \infty)=n_\infty=0.434$ is marked 
by a horizontal gray dashed line \cite{ninfNote}.
(b) Temperature dependence of the fluctuations of the mean density of gauge 
excitations $\delta n(T)$ and the derivative of the flux density 
$\Phi'(T)$ normalized to its minimum value at the cross over temperature
$\Phi'(T_m) = -15.16$. Vertical dashed gray line: location of $T_m \simeq 0.06 J_z$.
In addition, the specific heat is shown, 
obtained from ED and from the AGC methods, labeled accordingly.
For the AGC, $n(T)$ as fitted to the ED result is used. 
All ED data from an $N=6\times 6$ sites system. The AGC data
from an $N=20\times20$ system, with $N_R=20000$ realizations.       
}
\label{figThermodynamics}  
\end{figure}

To approximate the scale $T_R$, we evaluate $n(T)$ and its fluctuations 
$\delta n(T)$,  $\Phi(T)$ and its temperature derivative $\Phi'(T)$, 
as well as a thermodynamic observable, namely the specific heat $C_V$ {\it exactly} 
on a finite system of $N=6\times 6$ sites. We use
\begin{equation}
n(T) = \frac{1}{ZN} \mathrm{Tr}_\eta Z_{d(\eta)} n_\eta \,,
\label{eq:nT} 
\end{equation}
where $n_\eta$ is the number of gauge fields flipped off the uniform 
ground state sectors, excluding degenerate ground state sectors. The 
flux density is defined by  
\begin{equation}
\Phi(T) = \frac{1}{ZN} \sum_{\{\eta_\mathbf{r}\}} Z_{d(\eta)} 
\sum_\mathbf{r}  \Phi_\mathbf{r}\,, \quad 
\Phi_\mathbf{r} = \eta_{\mathbf{r}} \eta_{\mathbf{r} + 
\mathbf{e}_x - \mathbf{e}_y}.  
\label{eq:phiT} 
\end{equation}
First, Fig.~\ref{figThermodynamics}(a) shows that at very low temperatures, 
$n(T)$ and $\Phi(T)$ represent the gauge homogeneous ground state. Second, at
temperatures $T/J_z \sim 0.03$, well below the single gauge flip gap
$\Delta_1/J_z\simeq0.263$ \cite{Kitaev2006}, collective gauge excitations
lead to a rapid increase of $n(T)$, a downturn of $\Phi(T)$, and a region
of large fluctuations $\delta n(T)>n(T)$. Third, and for $T\gtrsim 0.1J_z
\equiv T_R$, the
system has essentially settled into a {\it completely random} gauge state
with its proper infinite temperature limiting value of 
$n_\infty\simeq 0.434$ for $N=6\times 6$ \cite{ninfNote}.  
In this regime the system can be considered as free fermions scattering 
from a fully random binary potential. 

In the crossover region $n(T)\sim1/[e^{(\Delta/T)^b}+1]$ with $\Delta =
0.06$, and $b = 3.5$ for the finite system \cite{noteCrossOver}. This
rather abrupt transition is likely due to gauge-gauge interactions and the
large degree of degeneracy of the gauge fields for a given number of fluxes
\cite{refAppDeg}. Considering the specific heat $C_V$ \cite{noteSpec} in
Fig.~\ref{figThermodynamics}(b), there is a clear release of entropy of the
fluxes or the gauges in the vicinity of $T_m\simeq 0.06 J_z$  
\cite{Motome2015b, noteFluxesEntropy}.

In view of $\delta n(T)/n(T)$ versus $T$, as in
Fig. \ref{figThermodynamics}(a,b), the AGC will work acceptably well for
$T\gtrsim T_R$. An indication of this is provided by evaluating $C_V$
within the AGC, using $n(T)$ as from the ED, averaging over $N_R=20000$
realizations, for a system with $N=20\times 20$. Obviously the agreement to
the exact result is excellent down to $T\approx T_R$, below which the AGC
does not account for all of the entropy release. To conclude: we confine
all subsequent AGC calculations to $T_R \lesssim T < \infty$, using a fully
random $\eta$ ensemble, i.e. $n(T)=0.5$

\begin{figure}
\includegraphics[width=0.95\columnwidth]{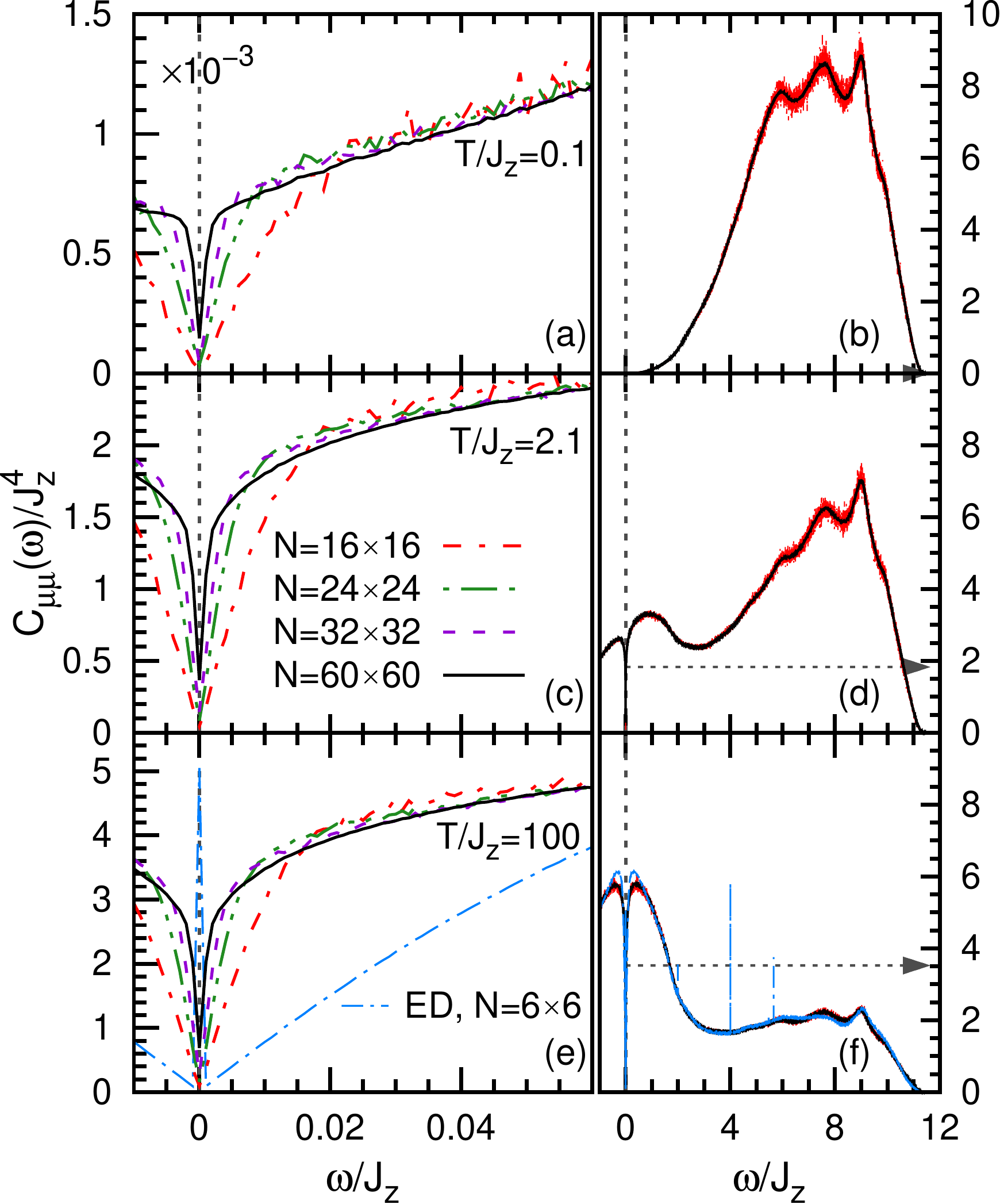}
\caption{$C_{\mu\mu}(\omega)$ obtained via the AGC method for various 
lattices $L=16-60$ and three different temperatures $T/J_z=0.1,2.1,100$ 
from top to bottom. The left panels show the low frequency behavior of 
$C_{\mu\mu}(\omega)$, emphasizing the development of the low frequency 
pseudogap with system size. The arrows in the right panels 
indicate the zero frequency extrapolation value from a 2nd order polynomial.
For $T/J_z=100$ the ED results for a system with $L=6$ are also displayed. 
}
\label{figCOmegaAGC}  
\end{figure}

In Fig.~\ref{figCOmegaAGC}, we present the energy current correlation
function obtained via the AGC method spanning three decades of temperature
$T/J=0.1,2.1,100$ and a binning of $\delta \omega =0.001$. The left panels 
of the plot highlight the low frequency behavior of $C_{\mu\mu}(\omega,T)$ 
while the right ones scan the complete positive frequency range.  
First, the qualitative and quantitative agreement between
the ED and the AGC method for all temperatures shown is remarkable 
\cite{refAppCom}. At high temperatures there is a low frequency hump, the 
weight of which reduces with temperature due to the occupation numbers of 
the matter fermions. At the same time, and since the sum rule does not 
change with temperature, more weight is accumulated at high 
frequencies, as in Fig.~\ref{figCOmegaED}.  

\begin{figure}
\includegraphics[width=0.95\columnwidth]{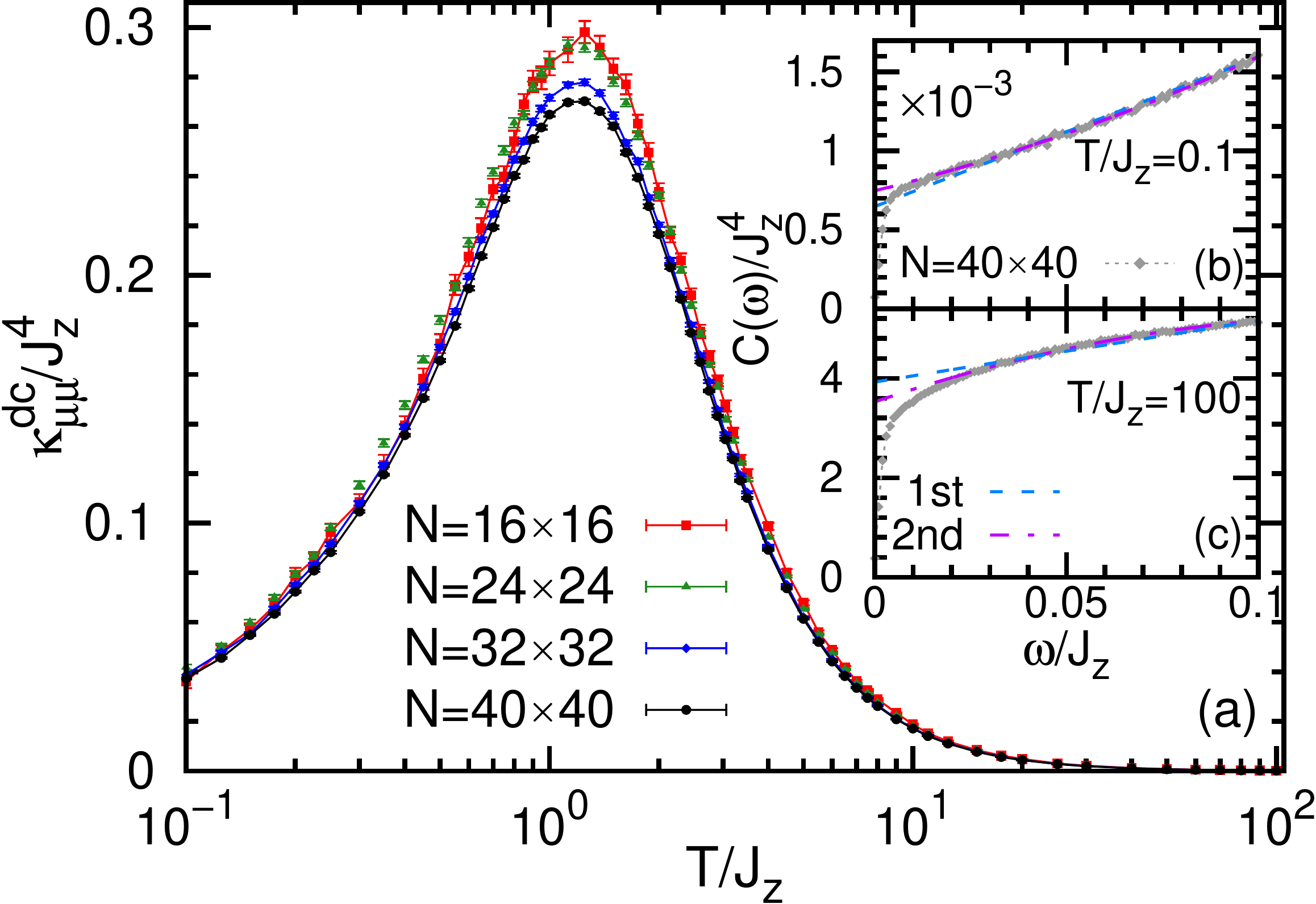}
\caption{(a) dc thermal conductivity $\kappa_{\mu\mu}^{dc}$ versus temperature 
for different system sizes,
obtained from fitting a second order polynomial to the low frequency 
regime of the correlation function. In the insets (b) and (c) this low 
frequency behavior of $C(\omega)$ for a system of $L=40$ at two 
temperatures $T/J_z = 0.1,100$ is shown, as well as,  
the corresponding linear and quadratic fit polynomials. The fitting range 
is restricted in the range $0.02\leq \omega/J_z \leq 0.12$ for 
all the fits.}\label{figKappaDC} 
\end{figure}

The left panels of Fig.~\ref{figCOmegaAGC} show, that apart from a smooth
downturn at $\omega/J_z\approx 1$ there is a second, sharp dip structure within
a very low energy range of $\omega/J_z\ll1$.  This narrow part of the
pseudogap displays a strong system size dependence, in stark contrast to the
rest of the frequency spectrum, for which larger system sizes merely render the
spectra smoother. It is interesting to note that the system sizes which can be
reached by ED do not display this low frequency structure,
Fig.~\ref{figCOmegaAGC}(e), rendering the use of the AGC method essential
\cite{refAppCom}.  This low-$\omega$ behavior with system size very much
suggests the pseudogap to close in the thermodynamic limit. Therefore we extract
a dc limit of the correlation function by fitting the data shortly before the
dip. Because of the slight curvature within the data, we choose to fit a
second order polynomial in the range $[0.02-0.12]$, incorporating 100 frequency
points, see also the inset of Fig.~\ref{figKappaDC} and its discussion.  The dc
limit extrapolation for all three temperatures is marked by the gray arrows at
the right panels of the plot. In addition, we note that we did not find any
Drude weights for the systems analyzed \cite{refAppCom}. In conclusion the
system will be a normal dissipative conductor in the thermodynamic limit.

Next, in Fig. \ref{figKappaDC}, we present the temperature dependence of
the dc thermal conductivity, Eq.~\eqref{eq:a6}, for different system sizes.
The overall behavior of $\kappa_{\mu\mu}^{dc}(T)$ versus $T$ is typical for spin
systems. It features a maximum at $T/J_z \approx 1$ and a $1/T^2$ decay at high
temperatures. To assess this result several sources of uncertainty have to
be mentioned. First, finite size effects are visible, which are however
satisfyingly small. Second, choosing a particular fit function and fitting
range induces an error. Its magnitude can be estimated from the two insets
Figs.~\ref{figKappaDC}(b), and \ref{figKappaDC}(c), where either a 1st ($c_0
+ c_1\omega$) or a 2nd order polynomial ($c_0 + c_1\omega+ c_2\omega^2$) is
used, leading to slightly different dc extrapolations. Similar variations
can be induced by changing the frequency window of the fit. In passing we
mention, that at high temperatures logarithmic fit functions, i.e. $\ln
(c_0 + c_1 \omega) $ also provide a good representation of $C(\omega\ll
1)$. Finally, the least square fit itself comprises an error which,  
however, is comparable with  the size of the symbols depicted.

\subsection{Fixed ground state gauge} \label{secUni}

As compared to the previous sections, fixing the gauge to a ground state
configuration, allows to obtain analytic expressions for the conductivity. While in
principle this only represents the limit $T/\Delta\rightarrow0$, it can nevertheless
be used to check the approach to low temperatures of the ED and AGC results. Moreover
it is instructive, to contrast a fictitious heat conductivity at \emph{all}
temperatures, arising from fixing the gauge to $\eta_{\mathbf{r}}=1$ with that
including the effects of thermally excited gauges. Since $\eta_{\mathbf{r}}=1$ is a
homogeneous state, we switch to momentum space, where the Hamiltonian and the current
can be written as
\begin{equation}
H=\sum_{\mathbf{k}}\mathbf{D}_{\mathbf{k}}^{\dagger}h_{\mathbf{k}}^{\phantom{\dagger}}
\mathbf{D}_{\mathbf{k}}^{\phantom{\dagger}}\,,\,\,\,{\cal J}_{\mu}=\sum_{\mathbf{k}}
\mathbf{D}_{\mathbf{k}}^{\dagger}L_{\mathbf{k},\mu}\mathbf{D}_{\mathbf{k}}^{
\phantom{\dagger}}\,,\label{a5}
\end{equation}
where boldface $ \mathbf{D}_{ \mathbf{k}}^{ \phantom{ \dagger}}=(d_{ \mathbf{k}}^{
\phantom{ \dagger}},d_{- \mathbf{k}}^{ \dagger})$ are `spinors', with $d_{
\mathbf{r}}^{ \dagger}= \sum_{ \mathbf{k}} \exp(-i \mathbf{k} \cdot \mathbf{r})d_{
\mathbf{k}}^{ \dagger}$.  We label the two entries by light symbols $D_{ \mathbf{k}
\alpha}^{ \phantom{ \dagger}}$, with $ \alpha=1,2$. Note that $D_{ \mathbf{k}
\alpha}^{ \phantom{ \dagger}}$ are destruction(creation) operators depending on $
\alpha=1$(2). Both, the Hamiltonian and current matrix elements, $h_{ \mathbf{k}}$ and
$L_{ \mathbf{k},x(y)}$ for the $x(y)$-directions are encoded in 2$ \times$2
matrices. From Eqs.~(\ref{eq:a2}), (\ref{eq:a3}), and  (\ref{eq:a10}) we get
\begin{eqnarray}
h_{\mathbf{k}}  &=&  \left[\begin{array}{cc}
e_{\mathbf{k}} & i\Delta_{\mathbf{k}}\\
i\Delta_{-\mathbf{k}} & -e_{-\mathbf{k}}
\end{array}\right]~, 
\quad 
L_{\mathbf{k},\mu}  =  l_{\mathbf{k},\mu}\left[\begin{array}{cc}
1 & 1\\
1 & 1
\end{array}\right]\label{eq:14}
\end{eqnarray} 
where, 
\begin{subequations} 
\begin{eqnarray} 
e_{\mathbf{k}} & = & 2[J_{z}-[J_{x}\cos(k_{x})+J_{y}\cos(k_{y})]~,\\
\Delta_{\mathbf{k}} & = & 2[J_{x}\sin(k_{x})+J_{y}\sin(k_{y})]~, \\
l_{\mathbf{k},\mu} & = & 2[J_{\mu}J_{z}\sin(k_{\mu})-J_{x}J_{y}
\sin(k_{\mu}-k_{\overline{\mu}})]\,,\hphantom{aaa} 
\end{eqnarray}
\end{subequations}
with $ \overline{ \mu}=y(x)$ for $ \mu=x(y)$. After Bogoliubov transformation
onto $h_{\mathbf{k}}$'s quasiparticle (QP) basis $c_{ \mathbf{k}}^{{(
\dagger)}}$, the Hamiltonian reads
\begin{equation} 
H =  \sum_{\mathbf{k}}\varepsilon_{\mathbf{k}}(c_{\mathbf{k}}^{\dagger}
c_{\mathbf{k}}^{\phantom{\dagger}}-\frac{1}{2})~, \quad \text{with} \quad 
\varepsilon_{\mathbf{k}}  =  \sqrt{e_{\mathbf{k}}^{2}+\Delta_{\mathbf{k}}^{2}}
\,.\label{eq:15}
\end{equation}
Remarkably, the current operator is \emph{invariant} under this Bogoliubov
transformation. I.e. also in the QP basis ${ \cal J}_{ \mu}= \sum_{ \mathbf{k}}
\mathbf{C}_{ \mathbf{k}}^{ \dagger}L_{ \mathbf{k}, \mu} \mathbf{C}_{ \mathbf{k}}^{
\phantom{ \dagger}}$, with $L_{ \mathbf{k}, \mu}$ identical to Eq. (\ref{eq:14}) and
$ \mathbf{C}_{ \mathbf{k}}^{ \phantom{ \dagger}}=(c_{ \mathbf{k}}^{ \phantom{
\dagger}},c_{- \mathbf{k}}^{ \dagger})$.  Since $L_{ \mathbf{k}, \mu}$ is not
diagonal, the energy current has both, QP and pairbreaking contributions. It is
satisfying to realize that $2\,l_{\mathbf{k},\mu}=$ $\varepsilon_{\mathbf{k}}\,
\partial\varepsilon_{\mathbf{k}}/\partial k_{\mu}$
Therefore, and because of Eqs.~(\ref{a5}), and (\ref{eq:14}), the naive 
expectation that the energy current can be written as
\begin{equation}
{\cal J}_{\mu}=\sum_{\mathbf{k}}\varepsilon_{\mathbf{k}}\frac{\partial\varepsilon_{
\mathbf{k}}}{\partial k_{\mu}}\,c_{\mathbf{k}}^{\dagger}c_{\mathbf{k}}^{\phantom{
\dagger}}+\textrm{pair breaking terms}\,,\label{a20}
\end{equation}
is indeed satisfied by our definition of the local energy density Eq. (\ref{eq:a3}).

Evaluating the current correlation function in the QP basis is straightforward.  We
get
\begin{eqnarray}
\lefteqn{C_{\mu\mu}^{0}(\omega)=\frac{2\pi}{N}\sum_{\mathbf{k}}\{2|l_{\mathbf{k},\mu}|^{2}[\,2f_{
\mathbf{k}}(1-f_{\mathbf{k}})\,\delta(\omega)+}\nonumber \\
 &  & \hphantom{aaa}f_{\mathbf{k}}^{2}\,\delta(\omega+2\varepsilon_{
\mathbf{k}})+(1-f_{\mathbf{k}})^{2}\,\delta(\omega-2\varepsilon_{\mathbf{k}})]\}
\,.\label{eq:21}
\end{eqnarray}
Where the superscript $0$ refers to the ground state gauge, the term $ \sim \delta(
\omega)$ represents the DW, and the remaining two addends are the
pairbreaking contributions

\begin{figure}[tb]
\includegraphics[width=0.95\columnwidth]{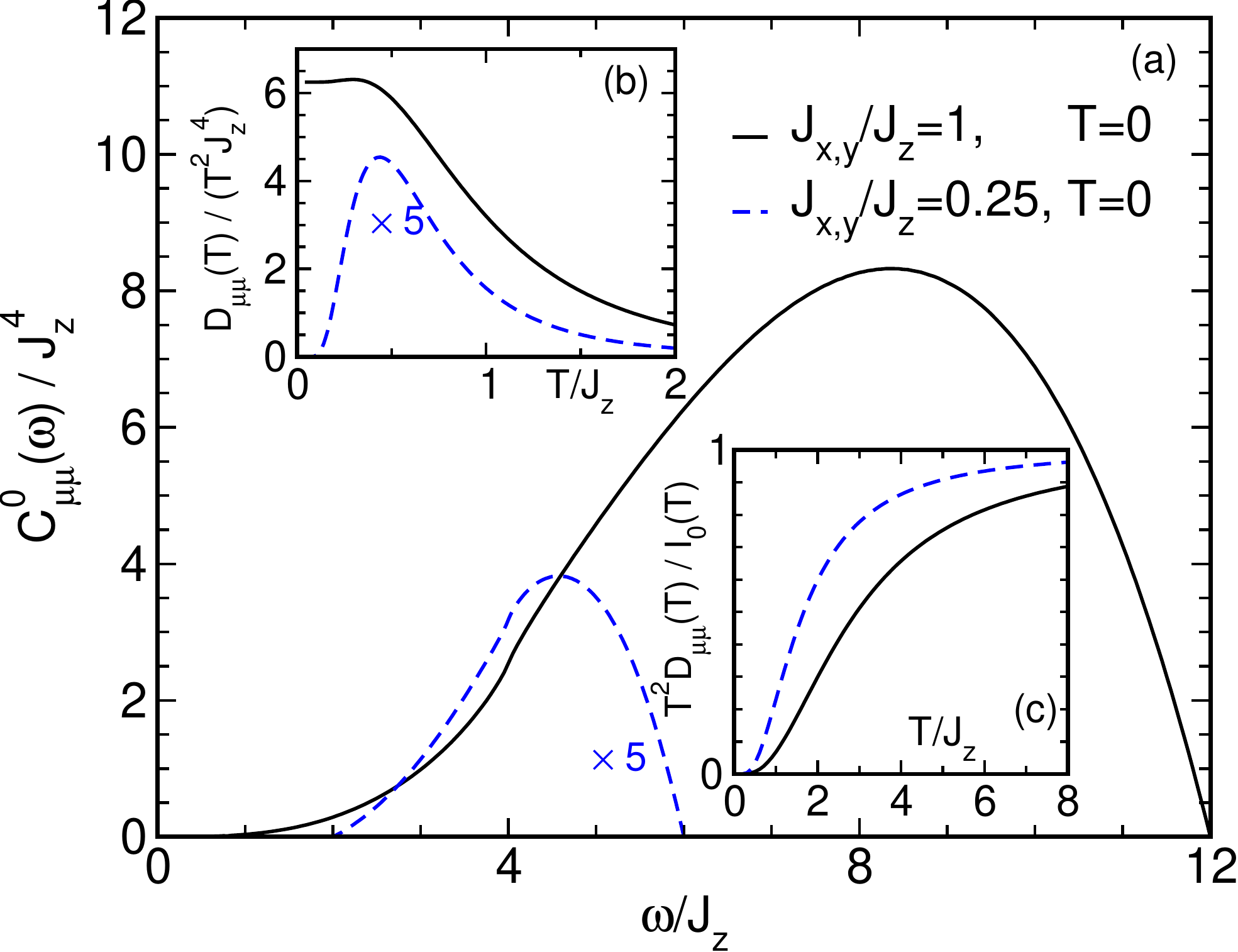}
\caption{(a) Black(blue) lines: $T{=}0$ regular part of dynamical current correlation
function $C_{\mu\mu}^{0}( \omega)$ versus frequency $ \omega{>}0$ using the ground
state gauge for gapless(ful) matter sector at $J_{x,y}/J_z=1(0.25)$.
Insets: (b) DW $D_{\mu\mu}/T^{2}$ versus temperature. (c) Ballistic weight
$T^{2}D_{\mu\mu}$ normalized to the weight of the regular part 
$I_{0}(T) = \dashint_{-\infty}^{ \infty}C_{\mu\mu}^{0}(\omega) \,d \omega$. \label{fig4}}
\end{figure}

Figure \ref{fig4}(a) shows $C_{\mu\mu}^{0}( \omega)$ for two representative
cases of $J_{x,y}/J_z=1$ ($J_{x,y}/J_z=0.25$) referring to a gapless
(gapped) matter sector. Several comment are in order. First, the regular
spectrum at small $\omega$ reflects the gap structure of the low energy
quasiparticle DOS combined with the energy current, leading to a power law
$C^0_{\mu\mu}(\omega) \propto \omega^{3}$ in the gapless case, while displaying a
linear onset above a finite gap. Second, within the spectrum a weak van-Hove
singularity arises from the saddlepoint of the dispersion
Eq.~(\ref{eq:15}). E.g.~for the gapless case in Fig.~\ref{fig4}, there is a log-singular derivative
of $C_{\mu\mu}^{0}( \omega)$ at $ \omega=4$, which is hardly noticeable on the
scale of the plot.  The inset Fig.~\ref{fig4}(b) depicts the Drude weight
divided by $T^{2}$ versus temperature. The main point is to demonstrate, that
in the gapless(gapped) case the Drude weight is \emph{finite} for any $T>0$ with
$D_{\mu\mu} \propto T^{2}$($\propto\exp(-a/T)$) for $T \ll1$.  This implies that
remaining within the ground state gauge, the system is a \emph{ballistic} energy
conductor, with infinite heat conductivity at any finite temperature. The inset
Fig.~\ref{fig4}(c) details another aspect of the DW, namely that the spectral
weight of the ballistic channel, i.e.~$T^{2}D_{\mu\mu}$, is of similar size than that of
the integrated regular spectrum $I_{0}(T)= \dashint_{-\infty}^{\infty}C_{\mu\mu}^{0}(
\omega) \,d \omega$.

\section{Conclusion}\label{secConcl}

In conclusion, we have studied the dynamical longitudinal heat transport of the
2D Kitaev model on the honeycomb lattice. Our conclusions are based on three 
complementary approaches, using the mapping of the spin Hamiltonian onto matter 
fermions and a $\mathbb{Z}_2$ gauge field. First, we 
have employed numerically exact diagonalization of small systems, up to 72
spin sites. Second, to reach system sizes of up to 7200 spin sites, we have
approximately restricted the complete gauge trace to only a random gauge
configuration, demonstrating that this leads to reliable results over a wide
range of temperatures. Finally, we have performed an analytical evaluation of
transport properties in the uniform gauge sector.

Among our main findings is, that fractionalization into Majorana matter and static 
gauge fields leaves a clear fingerprint on the spectrum of the current correlation 
function. In fact, thermally populated gauge excitations serve as an emergent disorder 
inducing an intrinsic energy scale for the relaxation of the matter fermion heat
currents. This relaxation leads to a clearly observable low-$\omega$
accumulation of spectral weight in the current correlation function, increasing
in intensity as the matter fermion density increases with temperature. We find
this low-$\omega$ spectral weight to display a zero frequency pseudogap, which
is strongly system size dependent. Based on finite size scaling, we have
concluded that in the thermodynamic limit, part of the pseudogap closes,
rendering the dc limit of the correlation function finite, albeit leaving a very
sharp low-$\omega$ depletion within the spectrum behind. Therefore we have shown
the 2D Kitaev model to be a normal dissipative heat conductor. This is in stark
contrast to the Kitaev ladder, which is an insulator with a vanishing Drude
weight and dc limit of the dynamical conductivity \cite{Metavitsiadis2016}, as
well as the one dimensional Kitaev chain, which is a ballistic conductor
\cite{Steinigeweg2016} and features a finite Drude weight (DW). We find, that
for the 2D Kitaev model, the DW is finite only on small systems or when gauge
excitations are completely neglected.

We caution that our finite size analysis cannot exclude extreme scenarios, in
which at system sizes way beyond our reach, the pseudogap ceases to close
and/or alters its variation with $\omega$, such as to remain with a zero
dc conductivity.

While current materials, proximate to the Kitaev model display a heat transport,
intricately intertwined with lattice degrees of freedom
\cite{Leahy2017,Hirobe2016,Hentrich2017}, it will be interesting to see if
future systems might show Majorana matter heat transport as depicted in
Fig.~\ref{figKappaDC}.

\section{Acknowledgments}
We thank C.~Hess, B.~B\"uchner, M.~Vojta, and S.~Rachel for fruitful
comments. Work of W.~B.~has been supported in part by the DFG through SFB 1143
and the NSF under Grant No.~NSF PHY11-25915. W.~B.~also acknowledges kind
hospitality of the PSM, Dresden.

{\it Note added}. -- After completion of this work, we have become aware of
related work by J. Nasu, J. Yoshitake, and Y. Motome, arXiv:1703.10395

\appendix
\numberwithin{equation}{section}

\section{Degeneracies}\label{appDeg} 

This appendix highlights the role of the degeneracies of the Kitaev model
with respect to the observable considered, namely the heat current
correlation function.  First, we recapitulate that the spectrum of the
Kitaev model is highly degenerate. This is due to the fact that a
`\emph{chain-flip}', i.e.~inverting the sign of $\eta_{\mathbf{r}}$ located
on \emph{all} $z$-links attached to any particular $xy$-chain is a unitary
transformation \cite{Feng2007}, leaving the fermionic spectrum invariant.
We emphasize, that this degeneracy is a physical property and unrelated to
spurious states which arise in some of the mappings \cite{Kitaev2006} from
spins to fermions and $\mathbb{Z}_{2}$ fields for the Kitaev model. While only the
case of open boundary conditions (OBCs) along the $xy$-chains is considered
in Ref.~[\citenum{Feng2007}], the degeneracy remains in place using
Eqs.~(\ref{eq:a2}), and (\ref{eq:a3}) with PBCs on the torus and also for
our square lattice geometry.

\begin{figure}[tb]
\centering{}\includegraphics[width=0.95\columnwidth]{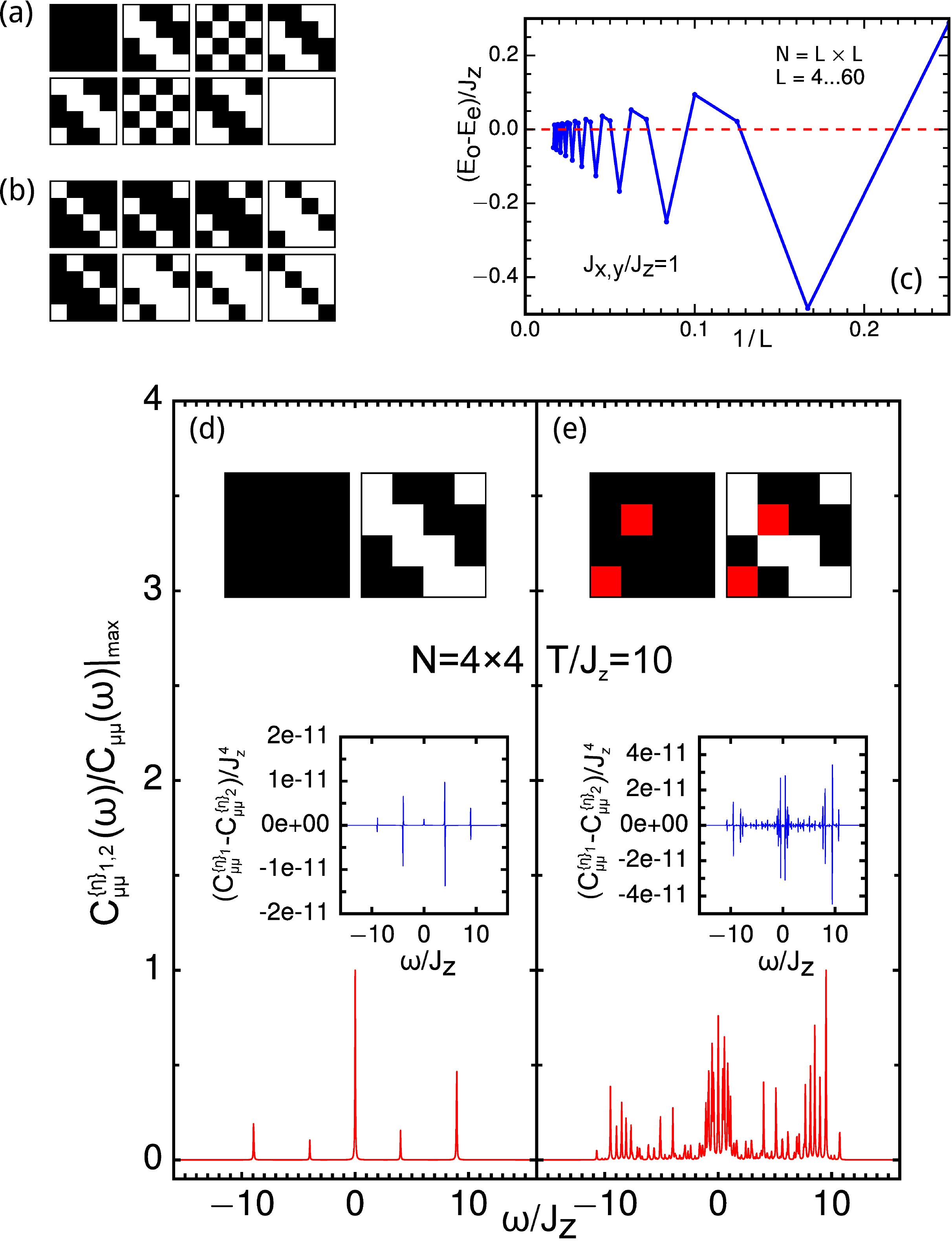}\caption{(a), 
(b): All $2^{L-1}$ gauge amplitudes for the gauge sectors $\{\eta^{0}\}_{e,o}$
of the two lowest fermionic ground state energies $E_{e}$, $E_{o}$
on a finite $L\times L=4\times4$ square lattice with PBCs. (a) {[}(b){]}
exhibit even{[}odd{]} number of line-flips. Complete fermionic spectrum
is degenerate within (a){[}(b){]}, respectively. (c): Collapse of
gauge sectors $\{\eta^{0}\}_{e,o}$ onto $2^{L}$ gauge sectors with
degenerate fermionic spectrum, versus $1/L=2k$, $k\in\mathbb{N}$
for $J_{x}=J_{y}=J_z$.  
$C_{\mu\mu}(\omega)$ on $L\times L=4\times4$, at $T/J_z=10$. (d) for two
degenerate ground state gauge configurations $\{\eta\}_{1,2}$, shown
in the graph, from the gauge sectors depicted in Fig. \ref{fig2}(a).
Inset: proves conductivities identical up to numerical error. (e)
for two fixed degenerate ground state gauge configurations $\{\eta\}_{1,2}$,
shown in the graph, from the gauge sectors depicted in Fig. \ref{fig2}(a),
including two flipped gauge fields as the red sites indicate. Inset:
proves conductivities identical up to numerical error.
\label{fig2}}
\end{figure}

On any finite system of $N=L\times L$ sites, and restricting to even $L=2k$
with $k\in\mathbb{N}$ hereafter, chain-flips will render each gauge
configuration $2^{L-1}$ fold degenerate. On these {\it finite} systems, the
ground state is either within the manifold of $\eta_{\mathbf{r}} = 1 \, \,
\forall\mathbf{r}$ \cite{Kitaev2006}, the \emph{homogeneous} sector, or in
a homogeneous sector, except for one `{\it line-flip}', i.e. with
$\eta_{\mathbf{r}}=1\,\,\forall\mathbf{r}$, except for a single ladder of
$z$-links with $\eta_{\mathbf{r}}=-1$. The energies $E_{e(o)}$ of these two
types of configurations are the two lowest of the system. For a $4\times4$
system these two gauge sectors are shown in
Fig.~\ref{fig2}(a),(b). Fig.~\ref{fig2}(c) shows the collaps of the ground
state energies for these two lowest energy sectors versus $L$. As
$L\rightarrow\infty$ this implies a $2^{L}$ fold degeneracy. Following the
same logic, all energies of the model are at least $2^{L}$-fold degenerate
in the thermodynamic limit.

The main point of this appendix is to exemplify, that not only the
Hamiltonian, but also the physical observable $C_{\mu\mu}(\omega)$ is
invariant under chain-flip operations.  In Fig. \ref{fig2}(d,e) we show
this for two sets of degenerate gauge configurations on a $4\times4$
system. Fig.~\ref{fig2}(d) depicts $C_{\mu\mu}(\omega)$, using the first
two gauge configurations from Fig.~\ref{fig2}(a), while Fig.~\ref{fig2}(e)
employs the latter two gauges, with however gauge fields flipped on the two
red sites.  Obviously $C_{\mu\mu}(\omega)$ is identical within each of the
two sets of gauges. Therefore, in the thermodynamic limit, the trace over
$\eta$ in Eq.~(\ref{eq:a11}) can in principle be restricted to one of the
$2^{L}$ identical subtraces over gauge sectors, which are
equivalent up to all chain-flip operations and a single line-flip.

\section{ED vs AGC for a small system}\label{appCom} 

\begin{figure}[tb]
\centering{}\includegraphics[width=.95\columnwidth]{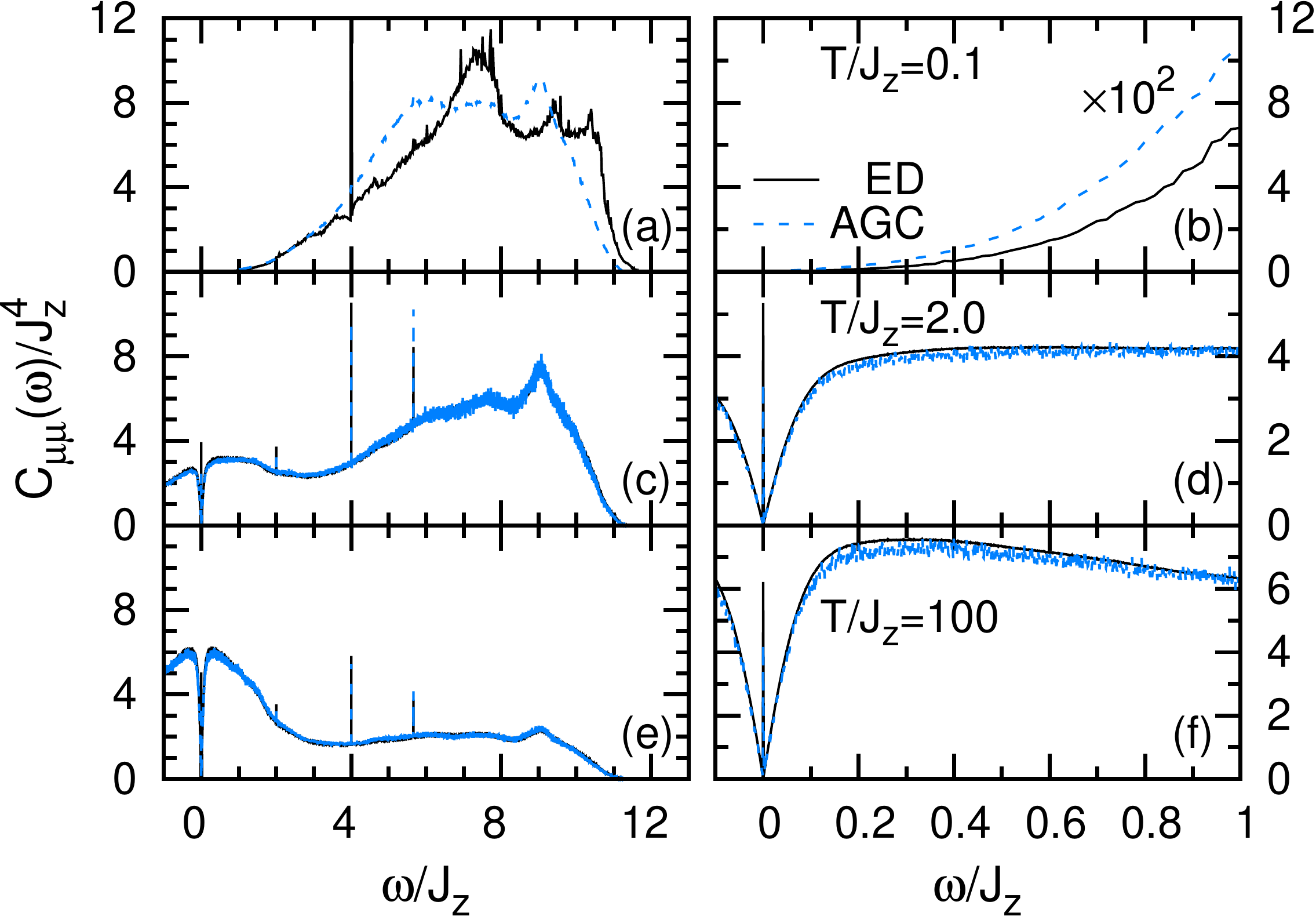}
\caption{Comparison of $C_{\mu\mu}(\omega)$ obtained via ED and AGC 
for a system of $N=6\times6$ sites, with $J_x=J_y=J_z$, at three temperatures 
$T/J_z=0.1,2.0,100$. The left panels display the full positive frequency 
spectrum of the correlation function at each temperature, while the right 
ones highlight the low frequency regime. 
The $\delta$-functions for the lowest temperature
are binned in windows of $\delta\omega=0.02$ while for the other two 
temperatures the bin size is $\delta\omega=0.001$. The AGC data are averaged 
over $N_R=50000$ random configurations with $n(T)=0.5$.
\label{figCompare}}
\end{figure}

In this appendix we make a direct comparison of the ED, Sec.~\ref{secED}, and 
the AGC, Sec.~\ref{secAGC}, methods for a small system of $N=6\times 6$ sites. 
In Fig.~\ref{figCompare}, we plot the correlation function 
$C_{\mu\mu}(\omega)$, obtained via the ED and the AGC methods at three 
different temperatures $T/J_z=0.1,2,100$ (from top to bottom). To anticipate 
finite size effects arising from the large mean level spacing of a system  
with a small linear dimension, we average over $N_R=50000$ 
random configurations with $n(T)=0.5$.  At high temperatures and down to 
$T\approx J_z$, the agreement between the two methods is impressive. 
Not only is the overall behavior of $C_{\mu\mu}(\omega)$ \emph{quantitatively} 
captured by the AGC method but also the fine structure yielded by singularities
 at the density of states. At the lowest temperature of $T=0.1J_z$, the AGC 
method clearly deviates from the exact high frequency structure of the 
correlation function, however the low frequency region $\omega < 4J_z$ is still
captured rather well.

Figure \ref{figCompare} fortifies the physical conclusions extracted in the main
text in the following two important aspects. First, it shows that the AGC method
is capable of detecting the existence of a finite Drude weight (DW), although it
fails to predict the correct weight of it. The latter does not come as a
surprise since the evaluation of the DW involves only degenerate states and the
AGC method is a random averaging approach.  Therefore, the absence of any trace
of a DW for the larger systems displayed in Fig.~\ref{figCOmegaAGC} shows that
the DW decays fast with system size, and the weight of the ballistic channel
completely disappears in the thermodynamic limit.  Second, Fig.~\ref{figCompare}
provides additional support for the notion of a closing of the low frequency dip
with system size extracted from the AGC. Namely, according to
Fig.~\ref{figCOmegaAGC}, ED displays only a very shallow pseudogap. However
Fig.~\ref{figCompare} proves, that this is not at variance with the AGC, but
solely due to the smaller system size of ED--which in turn we have overcome by
using the AGC method.


\begin{thebibliography}{99}
\bibitem{noteCrossOver}
In a free lattice gas picture where gauge flips are not interacting 
and they are gapped by $\Delta_1\simeq 0.263 J_z$, one would expect 
$n(T) \sim  1/(\exp(\Delta_1/T)+1)$.
\bibitem{noteSpec}
The specific heat is evaluated via
\begin{equation}
C_{V}=\langle C_{Vd}\rangle+\langle E_{m}^{2}\rangle
-\langle E_{m}\rangle^{2}\,. \label{eq:16} 
\end{equation}
$E_{m}=\sum_{i}(\varepsilon_{i}^{\eta}-\frac{1}{2})f_i$  
is the thermal energy of the matter fermions in a particular $\{\eta\}$
configuration, and $C_{Vd}=\sum_{i}\varepsilon_{i}^{\eta\,2}f_i(1-f_i)$   
their specific heat. Note that for only a single gauge field configuration, 
$C_{V}=C_{Vd}$, i.e. the two last terms in Eq. (\ref{eq:16}) refer to the 
contributions to $C_{V}$ from thermal gauge excitations.
\bibitem{noteFluxesEntropy}
The release of entropy $S(T)$ of the fluxes, or the gauge field respectively,  
can be readily seen from the qualitative agreement within the cross over 
region of the flux density temperature derivative with the specific heat 
$\Phi'(T) \propto C_V = T\partial S/\partial T$, 
Fig.~\ref{figThermodynamics}(b), and a simple partial integration.
\end{thebibliography}
\end{document}